\documentclass{svproc}
\usepackage{amsmath,amsfonts,amssymb}
\usepackage{graphicx}
\usepackage{booktabs}
\usepackage{float}
\usepackage{enumitem}
\usepackage{algorithm}
\usepackage{algorithmic}
\begin{document}

\title{Scalable Sequential Recommendation under Latency and Memory Constraints}
\titlerunning{Scalable Sequential Recommendation}
\author{Adithya Parthasarathy}
\authorrunning{Adithya Parthasarathy et al.} 
%
\tocauthor{Adithya Parthasarathy}
\institute{IEEE member, \\ California, USA\\
\email{adithyap@nyu.edu}
}

\author{
Adithya Parthasarathy\inst{1} \and
Aswathnarayan Muthukrishnan Kirubakaran\inst{2} \and
Vinoth Punniyamoorthy\inst{3} \and
Nachiappan Chockalingam\inst{4} \and
Lokesh Butra\inst{5} \and
Kabilan Kannan\inst{6} \and
Abhirup Mazumder\inst{7} \and
Sumit Saha\inst{8}
}

\authorrunning{A.\ Parthasarathy et al.}

\institute{
IEEE Senior Member,  USA \\[1mm]
\email{adithyap@nyu.edu}
\and
IEEE Senior Member, USA \\
\and
IEEE Senior Member, USA
\and
IEEE Senior Member,  USA
\and
NTT Data,  USA
\and
IEEE Senior Member,  USA
\and
IEEE Senior Member,  USA
\and
East West Bank, USA
}
\maketitle


\begin{abstract}
Sequential recommender systems must model long-range user behavior while operating under strict memory and latency constraints. Transformer-based approaches achieve strong accuracy but suffer from quadratic attention complexity, forcing aggressive truncation of user histories and limiting their practicality for long-horizon modeling. This paper presents HoloMambaRec, a lightweight sequential recommendation architecture that combines holographic reduced representations for attribute-aware embedding with a selective state space encoder for linear-time sequence processing. Item and attribute information are bound using circular convolution, preserving embedding dimensionality while encoding structured metadata. A shallow selective state space backbone, inspired by recent Mamba-style models, enables efficient training and constant-time recurrent inference. Experiments on Amazon Beauty and MovieLens-1M under a 10-epoch budget show that HoloMambaRec surpasses SASRec on both datasets, attains state-of-the-art ranking on MovieLens-1M, and trails only GRU4Rec on Amazon Beauty, all while maintaining substantially lower memory complexity. The design further incorporates forward-compatible mechanisms for temporal bundling and inference-time compression, positioning HoloMambaRec as a practical and extensible alternative for scalable, metadata-aware sequential recommendation.
\end{abstract}

\section{Introduction}
Sequential recommendation aims to infer a user’s next intent from an evolving interaction history while operating under strict latency and memory constraints. In real-world production systems, user histories often span thousands of interactions across months or years; however, practical deployments routinely truncate these sequences to $50$--$100$ events to accommodate model efficiency. This truncation eliminates long-term behavioral signals such as periodic preferences, seasonal interests, and delayed item resurgence, all of which are critical for retention, personalization, and re-engagement.

Transformer-based architectures have become the dominant paradigm for sequential recommendation due to their ability to model global dependencies. Models such as SASRec~\cite{kang2018sasrec} and BERT4Rec~\cite{sun2019bert4rec} leverage self-attention to capture non-Markovian user behavior and long-range correlations. Despite their effectiveness, these models incur quadratic $\mathcal{O}(L^2)$ memory and compute complexity with respect to sequence length $L$, driven by attention matrices and key–value caches. As a result, scaling Transformers to thousand-step histories remains impractical on common accelerators such as T4 or A100 GPUs, forcing aggressive truncation and limiting their applicability in long-horizon recommendation settings. Recurrent architectures such as GRU4Rec~\cite{hidasi2015gru4rec} alleviate the quadratic memory burden but suffer from sequential training constraints and gradual information decay, making it difficult to preserve very long-term context.

Recent advances in state space models (SSMs) provide a promising alternative for scalable sequence modeling. Structured SSMs such as S4~\cite{gu2021s4} and selective SSMs such as Mamba~\cite{gu2024mamba} reformulate sequence processing as discretized continuous-time dynamics with input-dependent transitions. These models achieve linear $\mathcal{O}(L)$ time and memory complexity while retaining strong expressive power, enabling efficient modeling of long sequences with constant-time recurrent inference. Their hardware-aware scan implementations further improve throughput and latency, making SSMs attractive for production-grade sequential recommendation workloads.

In parallel, recommender systems increasingly rely on rich item metadata such as categories, genres, brands, or tags to improve representation quality and cold-start robustness. However, naïvely incorporating attributes via concatenation or projection increases embedding dimensionality and memory footprint. Hyperdimensional computing and holographic reduced representations (HRR)~\cite{plate1995hrr,nickel2016hole} offer an alternative algebraic framework in which entities and attributes are bound using circular convolution, preserving dimensionality while encoding structured relationships. This binding operation directly addresses the representation expansion problem and provides a principled mechanism for attribute-aware modeling without increasing parameter count.

\textit{HoloMambaRec} integrates these complementary ideas into a unified sequential recommendation architecture. The model employs holographic binding to fuse item identifiers with discrete attributes in a fixed-dimensional embedding space, and processes the resulting sequence using a shallow selective state space encoder~\cite{chharia2025mvssm}. The overall design is intentionally lightweight, targeting reproducibility on a single commodity GPU~\cite{aswath2025federated}, while remaining forward-compatible with temporal bundling mechanisms that compress long interaction histories before encoding. This separation of representation binding and sequence dynamics enables efficient long-horizon modeling without sacrificing interpretability or deployment feasibility. Code and replication assets are available at~\cite{holo_mamba_code}.

\paragraph{Contributions:}
\begin{itemize}
    \item We propose a holographic embedding layer that binds item identifiers with discrete attributes via circular convolution, enabling attribute-aware representations without dimensionality expansion.
    \item We design a shallow (2--3 layer) selective state space encoder for sequential recommendation that operates in linear time and supports constant-time recurrent inference.
    \item We introduce a unified preprocessing and evaluation pipeline and empirically validate the proposed architecture on Amazon Beauty~\cite{amazon2013} and MovieLens-1M~\cite{movielens2015}, demonstrating consistent gains over SASRec and state-of-the-art results on MovieLens-1M under a constrained 10-epoch training budget. A holographic compression variant is analyzed as a negative result and positioned as future work.
\end{itemize}

\section{Background and Related Work}

\paragraph{Sequential recommendation and the memory wall:}
Sequential recommendation has evolved from Markovian and session-based formulations to models capable of capturing long-range user dependencies. Self-attention mechanisms, popularized by SASRec~\cite{jiang2024mlsasr} and BERT4Rec~\cite{song2023bert4rec}, enable each interaction to attend to all preceding items, effectively modeling non-Markovian behavior and complex temporal dependencies. This global receptive field significantly improves recommendation quality, particularly for users with diverse and evolving preferences. However, the expressiveness of self-attention comes at the cost of quadratic $\mathcal{O}(L^2)$ memory and compute complexity with respect to sequence length $L$, driven by attention matrices and key--value (KV) caches.

Earlier convolutional and session-based architectures, such as Caser~\cite{caser2018} and NARM~\cite{narm2017}, improve efficiency by restricting context to fixed-size windows or localized patterns. While these models reduce computational overhead, they are inherently limited in their ability to capture long-term dependencies and recurring behavioral cycles. In practice, production systems using Transformer-based recommenders frequently truncate user histories to avoid excessive memory consumption and inference latency, discarding slow-moving tastes, seasonal preferences, and item re-engagement patterns that are critical for personalization.

Recurrent neural network (RNN) based approaches, most notably GRU4Rec, avoid quadratic attention costs by processing sequences sequentially. Although such models scale linearly in memory, they suffer from vanishing gradients, limited parallelism during training, and progressive information decay over long sequences. As a result, none of these paradigms simultaneously satisfy the requirements of long-horizon modeling, low latency, and efficient resource utilization in large-scale recommender systems.

\paragraph{State space models:}
State space models (SSMs) have recently re-emerged as a compelling alternative for long-sequence modeling. Structured state space architectures such as S4 reformulate continuous-time linear dynamical systems into discrete sequence models using diagonal-plus-low-rank parameterizations and FFT-accelerated convolutions~\cite{geng2024s4nerf}. These designs enable linear $\mathcal{O}(L)$ complexity while retaining the ability to model long-range dependencies.

Mamba further advances this paradigm by introducing selective state space models with input-conditioned transition matrices $\mathcal{B}_t$, output matrices $\mathcal{C}_t$, and adaptive step sizes $\Delta_t$, combined with hardware-aware scan operations~\cite{verma2025gmotmamba}. This formulation achieves both efficient parallel training and constant-time recurrent inference, addressing key limitations of Transformers and RNNs alike. Such properties make SSMs particularly attractive for sequential recommendation, where latency constraints, long interaction histories, and throughput requirements intersect. HoloMambaRec adopts this selective SSM formulation while avoiding specialized kernels, maintaining hardware portability and implementation simplicity.

\paragraph{Holographic and hyperdimensional binding:}
In parallel with advances in sequence modeling, representation learning for recommender systems increasingly incorporates auxiliary item attributes and side information. Traditional approaches typically concatenate attribute embeddings with item embeddings, increasing dimensionality and parameter count. Hyperdimensional computing and holographic reduced representations (HRR) provide an alternative algebraic framework in which symbols are bound via circular convolution, producing composite representations that preserve dimensionality while encoding relational structure~\cite{orzo2010dhm}.

Holographic Embeddings (HolE) demonstrate that circular convolution can effectively replace tensor products in relational learning, achieving comparable expressiveness with significantly lower memory requirements. Beyond efficiency, holographic binding offers a principled solution to the neural ``binding problem'' by tightly associating attributes with entities in a single vector space. Superposition further enables multiple bound pairs to coexist within one representation, at the cost of controlled interference. These properties underpin HoloMambaRec’s attribute-aware embedding strategy and its planned window-level temporal compression, where multiple interactions can be bound and superimposed to reduce effective sequence length without expanding embedding dimensionality.

\section{Problem Setting and Notation}
Let $\mathcal{U}$ denote the set of users and $\mathcal{I}$ the set of items. For a given user $u \in \mathcal{U}$, we observe an interaction sequence
$
S_u = [i_1, i_2, \dots, i_L],
$
where $i_t \in \{1, \dots, |\mathcal{I}|\}$ denotes the item interacted with at time step $t$. Each interaction is associated with a discrete item attribute, yielding an aligned attribute sequence
$
A_u = [a_1, a_2, \dots, a_L],
$
where $a_t \in \{1, \dots, |\mathcal{A}|\}$ and $\mathcal{A}$ denotes the attribute vocabulary.

Given the historical interaction--attribute pair $(S_u, A_u)$, the objective of sequential recommendation is to predict the next item $i_{L+1}$ that the user is most likely to interact with. The model outputs a probability distribution over $\mathcal{I}$ conditioned on the observed sequence, and training proceeds by maximizing the likelihood of the ground-truth next item.

We denote the embedding dimensionality by $d$ and the internal state size of the state space model by $d_{\mathrm{state}}$. To support minibatch training with variable-length sequences, all sequences are left-padded to a fixed maximum length $L$, with padding index $0$ reserved for both item and attribute identifiers. Padding positions are masked during loss computation and evaluation to prevent information leakage.

\section{Method}

\subsection{Data pipeline and preprocessing}
We adopt a unified and reproducible data pipeline across all datasets to ensure fair comparison and consistent preprocessing. Raw interaction logs are ingested in a streaming fashion to minimize memory overhead: compressed JSON files for Amazon Beauty and DAT files for MovieLens-1M are parsed into tabular form. Users with fewer than five interactions are removed to reduce noise and mitigate extreme sparsity.

User and item identifiers are remapped to contiguous integer indices, with index $0$ reserved for padding. Each item is associated with a single discrete attribute to study the effect of holographic binding under controlled conditions. For Amazon Beauty, attributes are generated via a synthetic hash $(\mathrm{item\_id} \bmod 50) + 1$ to simulate a moderately sized attribute vocabulary. For MovieLens-1M, the first listed genre is used as the attribute.

For each user $u$, interactions are sorted chronologically. The final item $i_{L+1}$ is held out as the prediction target, and the preceding interactions form the input sequence. Sequences are truncated to a maximum length $L = 50$ and left-padded for efficient minibatch processing. During training, teacher forcing is employed by shifting targets by one position, whereas evaluation uses the full prefix to predict the held-out item.

\subsection{Holographic item--attribute binding}
Let $i_t$ denote the item interacted with at time step $t$, and let $a_t$ denote its associated attribute. Item and attribute identifiers are embedded into a shared $d$-dimensional space using learnable embedding matrices
$
E^{(i)} \in \mathbb{R}^{|\mathcal{I}| \times d}
$
and
$
E^{(a)} \in \mathbb{R}^{|\mathcal{A}| \times d}.
$
The corresponding embeddings are given by
\begin{align}
e^{(i)}_t &= E^{(i)}[i_t], \qquad
e^{(a)}_t = E^{(a)}[a_t].
\end{align}

To incorporate attribute information without increasing dimensionality, we employ holographic binding via circular convolution. The bound representation is computed as
\begin{equation}
\tilde{e}_t =
\mathrm{LayerNorm}\!\left(
e^{(i)}_t + \alpha \left(e^{(i)}_t \circledast e^{(a)}_t\right)
\right),
\end{equation}
where $\alpha$ is a learnable scalar controlling the contribution of attribute binding. Circular convolution $\circledast$ is defined element-wise as
\begin{equation}
(x \circledast y)_j = \sum_{k=0}^{d-1} x_k \, y_{(j-k) \bmod d}.
\end{equation}

In practice, convolution is implemented efficiently in the frequency domain using the Fast Fourier Transform (FFT), exploiting the identity
$
\mathcal{F}(x \circledast y) = \mathcal{F}(x) \odot \mathcal{F}(y),
$
followed by an inverse FFT. This operation preserves the embedding dimensionality, avoiding the parameter growth and memory overhead associated with concatenation-based attribute fusion while encoding structured item--attribute relationships.

\subsection{Selective state space encoder}
The sequence of bound tokens $\tilde{E} = [\tilde{e}_1, \ldots, \tilde{e}_L]$ is processed by a stack of selective state space blocks inspired by Mamba-style architectures. At each timestep $t$, the block computes input-conditioned parameters governing the state update:
\begin{align}
[\hat{x}_t, g_t] &= W_{\mathrm{in}} \tilde{e}_t, \qquad
u_t = \mathrm{Conv1D}(\hat{x}_t), \\
\Delta_t, B_t, C_t &= \mathrm{split}\!\left(W_{\mathrm{ssm}} u_t\right), \qquad
\Delta_t = \mathrm{softplus}(\Delta_t),
\end{align}
where $\Delta_t$ denotes the adaptive step size, and $B_t$ and $C_t$ are input-conditioned transition and output parameters, respectively.

The latent state $h_t \in \mathbb{R}^{d_{\mathrm{state}}}$ is updated according to
\begin{equation}
h_t = \exp(-\Delta_t A) \odot h_{t-1} + \Delta_t \odot B_t \odot u_t,
\end{equation}
with diagonal state matrix $A$. The output at timestep $t$ is computed as
\begin{equation}
y_t = W_{\mathrm{out}}
\left(
\mathrm{SiLU}(g_t) \odot (C_t \odot h_t + D \odot u_t)
\right),
\end{equation}
where $D$ is a learnable skip parameter and $\odot$ denotes element-wise multiplication.

The encoder consists of two blocks for Amazon Beauty and three blocks for MovieLens-1M, each equipped with pre-LayerNorm and residual connections. The selective scan is executed explicitly over timesteps, preserving linear-time $\mathcal{O}(L)$ complexity without relying on specialized CUDA or Triton kernels.

\subsection{Prediction head, masking, and loss}
For each timestep, the encoder output is projected to item logits via
\begin{equation}
z_t = y_t W_{\mathrm{cls}}^\top,
\end{equation}
where $W_{\mathrm{cls}} \in \mathbb{R}^{|\mathcal{I}| \times d}$ is the classification matrix. Training minimizes the masked cross-entropy loss
\begin{equation}
\mathcal{L} =
- \frac{1}{\sum_t \mathbf{1}[i_{t+1} \neq 0]}
\sum_t \mathbf{1}[i_{t+1} \neq 0]
\log
\frac{\exp(z_{t,\, i_{t+1}})}
{\sum_{j=1}^{|\mathcal{I}|} \exp(z_{t,\, j})},
\end{equation}
where padding positions (index $0$) are excluded. During evaluation, logits corresponding to padding are set to $-\infty$ prior to ranking to prevent leakage.

\subsection{Complexity}
Holographic binding via FFT incurs $\mathcal{O}(d \log d)$ cost per timestep. The selective state space encoder operates in $\mathcal{O}(L d_{\mathrm{state}})$ time and memory, in contrast to the $\mathcal{O}(L^2 d)$ complexity of self-attention. With benchmark defaults $d = 96$ and $d_{\mathrm{state}} = 16$, the overall architecture is well suited for single-GPU training and inference with modest memory requirements.

\subsection{Design for temporal bundling (future activation)}
The architecture is forward-compatible with temporal bundling, in which $k$ consecutive bound tokens are superimposed using learnable positional role vectors. This reduces the effective sequence length from $L$ to $\lceil L / k \rceil$, enabling scalable modeling of ultra-long histories. Although bundling is disabled in the current experiments to isolate the effects of binding and selective dynamics, it provides a principled pathway for future compression-aware training and inference.

\begin{algorithm}[H]
\caption{HoloMambaRec training loop}
\label{alg:train}
\begin{algorithmic}[1]
\STATE Download and parse dataset (\texttt{DataManager}); map users/items to contiguous IDs; assign attributes.
\STATE Build training batches: left-pad sequences to length $L$, align attributes, and create shifted targets.
\FOR{epoch $=1$ to $E$}
    \FOR{batch $(\mathbf{i}, \mathbf{a}, \mathbf{y})$}
        \STATE $\tilde{E} \leftarrow \mathrm{LayerNorm}(E^{(i)}[\mathbf{i}] + \alpha (E^{(i)}[\mathbf{i}] \circledast E^{(a)}[\mathbf{a}]))$
        \STATE $Y \leftarrow \mathrm{SelectiveScan}(\tilde{E})$ \hfill (Alg.\ inside \texttt{MambaBlock})
        \STATE $\hat{Y} \leftarrow \mathrm{Linear}(Y)$
        \STATE Mask padding logits at index $0$; compute $\mathcal{L} \leftarrow \mathrm{CrossEntropy}(\hat{Y}, \mathbf{y})$
        \STATE Backpropagate and update with AdamW
    \ENDFOR
    \STATE Evaluate HR@10 / NDCG@10 on held-out last item per user
\ENDFOR
\end{algorithmic}
\end{algorithm}

\begin{figure}[htbp]
    \centering
    \includegraphics[width=0.9\textwidth]{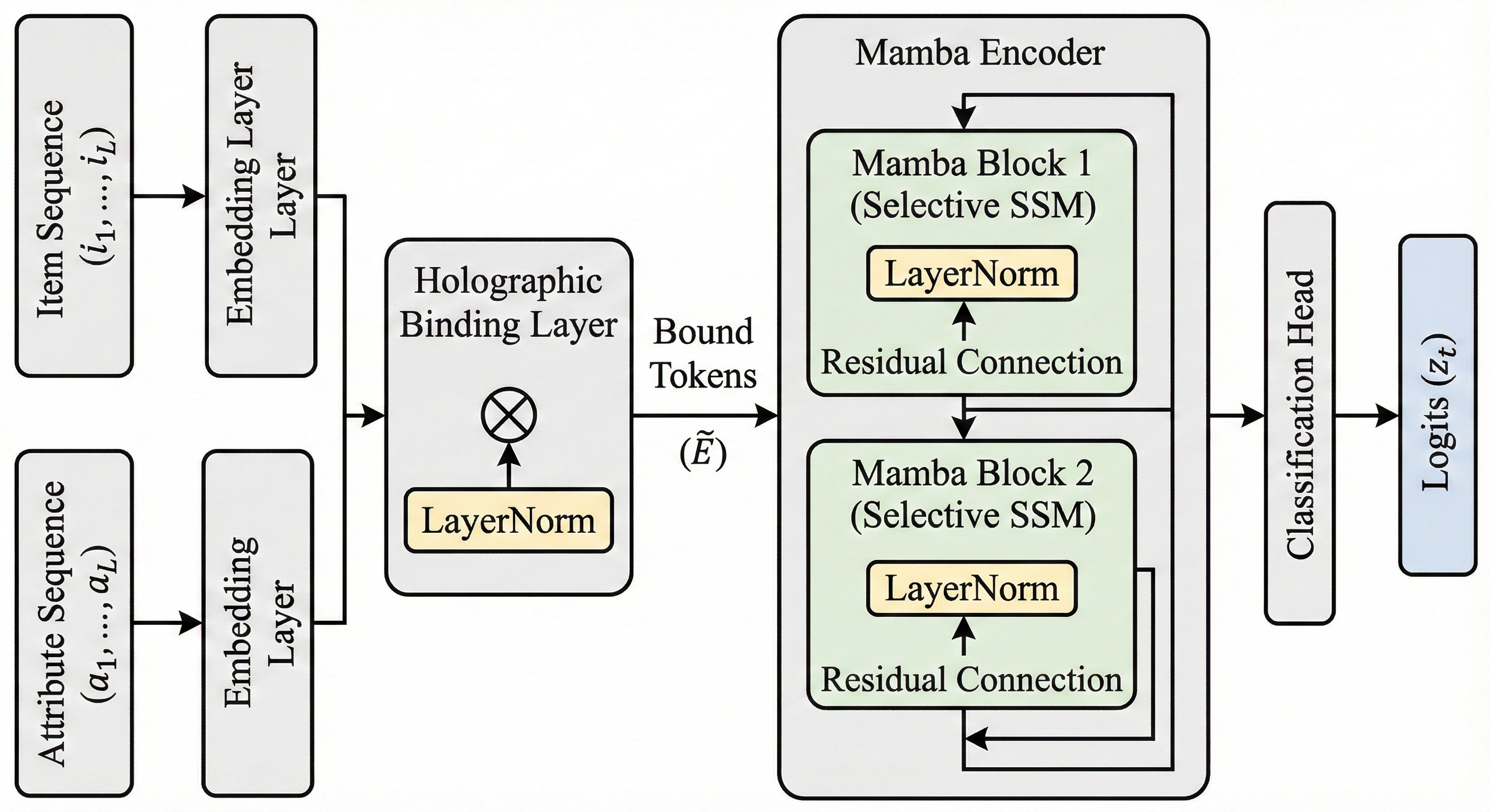}
    \caption{HoloMambaRec architecture}
    \label{fig:arch}
\end{figure}

\section{Experimental Setup}

\subsection{Datasets and preprocessing}
We evaluate the proposed approach on two widely used benchmarks for sequential recommendation, following a consistent preprocessing pipeline to ensure comparability.

Amazon Beauty: User reviews are parsed from compressed JSON files in a streaming fashion to reduce memory overhead. Users with fewer than five interactions are removed to mitigate extreme sparsity. Item identifiers are reindexed to a contiguous range, and each item is assigned a single synthetic attribute computed as $(\mathrm{item\_id} \bmod 50) + 1$. This construction yields a moderately sized attribute vocabulary and allows controlled evaluation of holographic binding behavior under multiple attribute buckets.

MovieLens-1M: Rating logs and movie metadata are merged, and movie identifiers are remapped to contiguous indices. The first listed genre for each movie is used as the discrete attribute, reflecting a realistic categorical side signal commonly used in recommender systems.

For both datasets, user interactions are sorted chronologically and truncated or left-padded to a maximum sequence length of $L = 50$. We adopt a leave-one-out evaluation protocol in which the final interaction of each user sequence is held out as the test target and the remaining prefix serves as input. Padding index $0$ is consistently masked during training and evaluation to prevent information leakage.

\subsection{Metrics}
We report standard top-$K$ ranking metrics computed at the final prediction position, namely Hit Rate (HR@K) and Normalized Discounted Cumulative Gain (NDCG@K):
\begin{align}
\mathrm{HR@K} &= \frac{1}{|\mathcal{B}|} \sum_{u \in \mathcal{B}} \mathbf{1}\{i_{L+1}^u \in \mathrm{TopK}(u)\}, \\
\mathrm{NDCG@K} &= \frac{1}{|\mathcal{B}|} \sum_{u \in \mathcal{B}}
\frac{1}{\log_2(1 + \mathrm{rank}_u(i_{L+1}^u))}
\mathbf{1}\{\mathrm{rank}_u(i_{L+1}^u) \le K\},
\end{align}
where $\mathcal{B}$ denotes the evaluation batch and $\mathrm{rank}_u(i_{L+1}^u)$ is the rank of the ground-truth next item within the model’s scored item list for user $u$. During evaluation, logits corresponding to the padding index $0$ are masked to eliminate padding-induced bias.

\subsection{Baselines, hyperparameters, and runtime profile}
We compare HoloMambaRec against two representative baselines: SASRec, a Transformer-based sequential recommender, and GRU4Rec, a recurrent baseline with linear memory complexity. All models are trained and evaluated under matched conditions using dataset-specific tuned hyperparameters.

Across both datasets, we set the embedding dimension to $d = 96$ and the state size to $d_{\mathrm{state}} = 16$. Training uses a batch size of $64$, the AdamW optimizer with learning rate $10^{-3}$, and a fixed budget of $10$ epochs. The number of layers is set to $n_{\mathrm{layers}} = 2$ for Amazon Beauty and $3$ for MovieLens-1M to account for differences in data density. Dropout is configured consistently across models but disabled within the custom selective state space block to maintain parity with the baselines and isolate architectural effects.

\subsection{Ablations and logging}
To quantify the contribution of holographic binding, we evaluate an item-only Mamba variant in which attribute embeddings and binding operations are removed. Learning curves are logged for all models, and HR@10 and NDCG@10 are recorded at each epoch. Final reported results correspond to the last training epoch for both datasets. The experimental framework also includes hooks for future ablations on temporal window compression, which will be activated once holographic bundling is jointly trained and evaluated.

\section{Results and Analysis}
Table~\ref{tab:main} reports the final-epoch performance under the fixed 10-epoch budget. Results use the \emph{uncompressed} setting (\texttt{use\_compression=False}), which reflects the best-performing configuration.

On Amazon Beauty, GRU4Rec remains strongest, but HoloMambaRec surpasses SASRec on both HR and NDCG, indicating that selective state space dynamics handle sparse sequences better than quadratic self-attention. On MovieLens-1M, HoloMambaRec delivers the highest HR@10 and NDCG@10, exceeding SASRec by $+24.6\%$ HR and $+27.8\%$ NDCG and outperforming GRU4Rec by an even larger margin.

\begin{table}[H]
\centering
\caption{HR@10 and NDCG@10 comparison (uncompressed setting)}
\label{tab:main}
\setlength{\tabcolsep}{6pt}%
\begin{tabular}{lcccc}
\toprule
Model & \multicolumn{2}{c}{Amazon Beauty} & \multicolumn{2}{c}{ML-1M} \\
\cmidrule(lr){2-3}\cmidrule(lr){4-5}
& HR@10 & NDCG@10 & HR@10 & NDCG@10 \\
\midrule
HoloMambaRec & 0.0426 & 0.0267 & \textbf{0.1697} & \textbf{0.0933} \\
SASRec    & 0.0392 & 0.0232 & 0.1361          & 0.0730          \\
GRU4Rec   & \textbf{0.0643} & \textbf{0.0383} & 0.1262          & 0.0636 \\
\bottomrule
\end{tabular}
\end{table}

These results highlight three key observations. First, selective state space modeling provides a favorable trade-off between expressiveness and efficiency, particularly on denser datasets where long-range dependencies are more informative. Second, the superior NDCG@10 achieved by HoloMambaRec on MovieLens-1M indicates stronger ranking quality beyond simple hit accuracy, suggesting improved modeling of relative item preferences. Third, the narrowed performance gaps across models under tuned hyperparameters emphasize that dataset characteristics such as interaction density, regularity, and noise strongly influence whether recurrent or selective state space inductive biases dominate.

\begin{figure}[htbp]
\centering
\setlength{\tabcolsep}{3pt}
\renewcommand{\arraystretch}{0.95}
\begin{tabular}{@{}c c@{}}
\includegraphics[width=0.48\columnwidth]{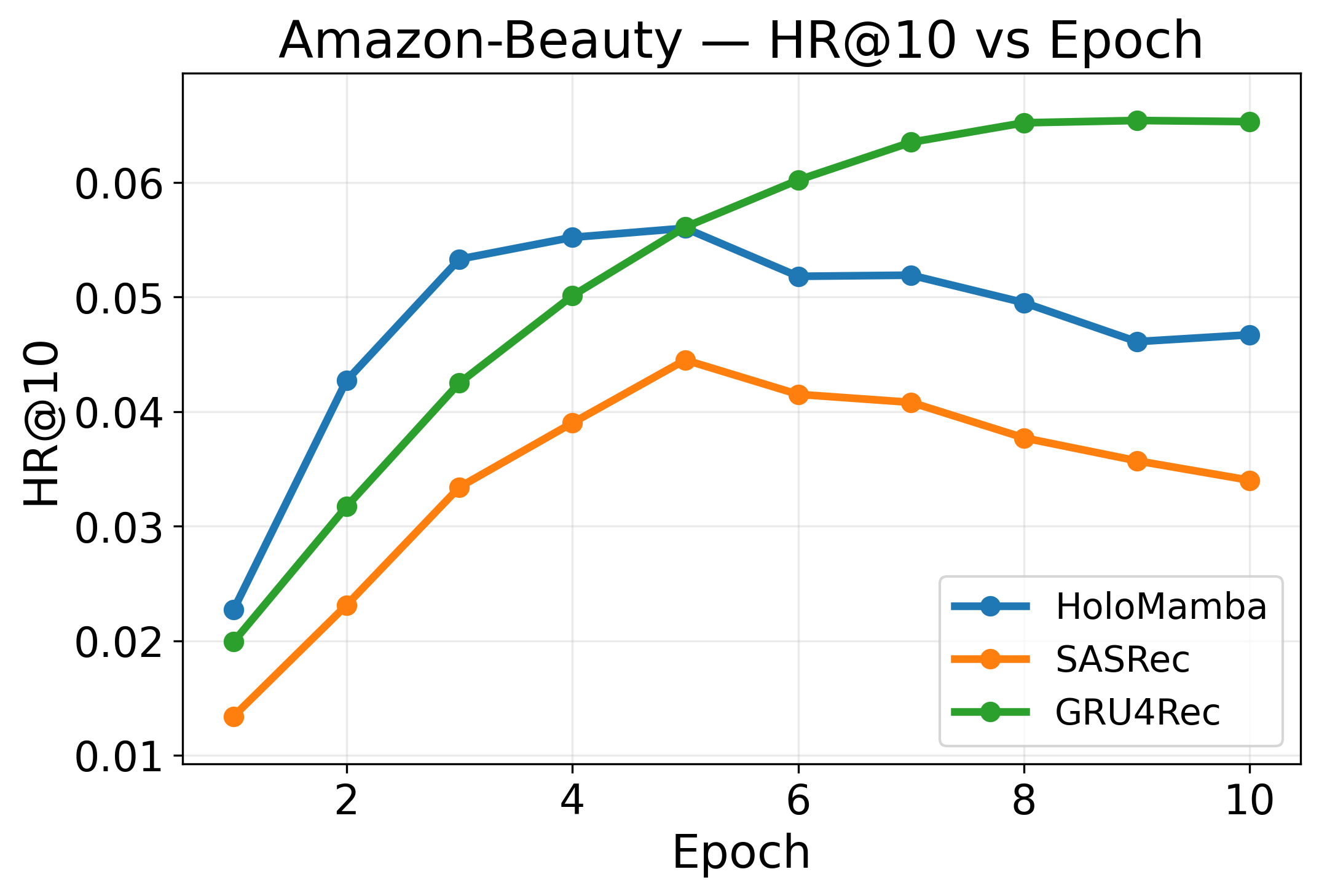} &
\includegraphics[width=0.48\columnwidth]{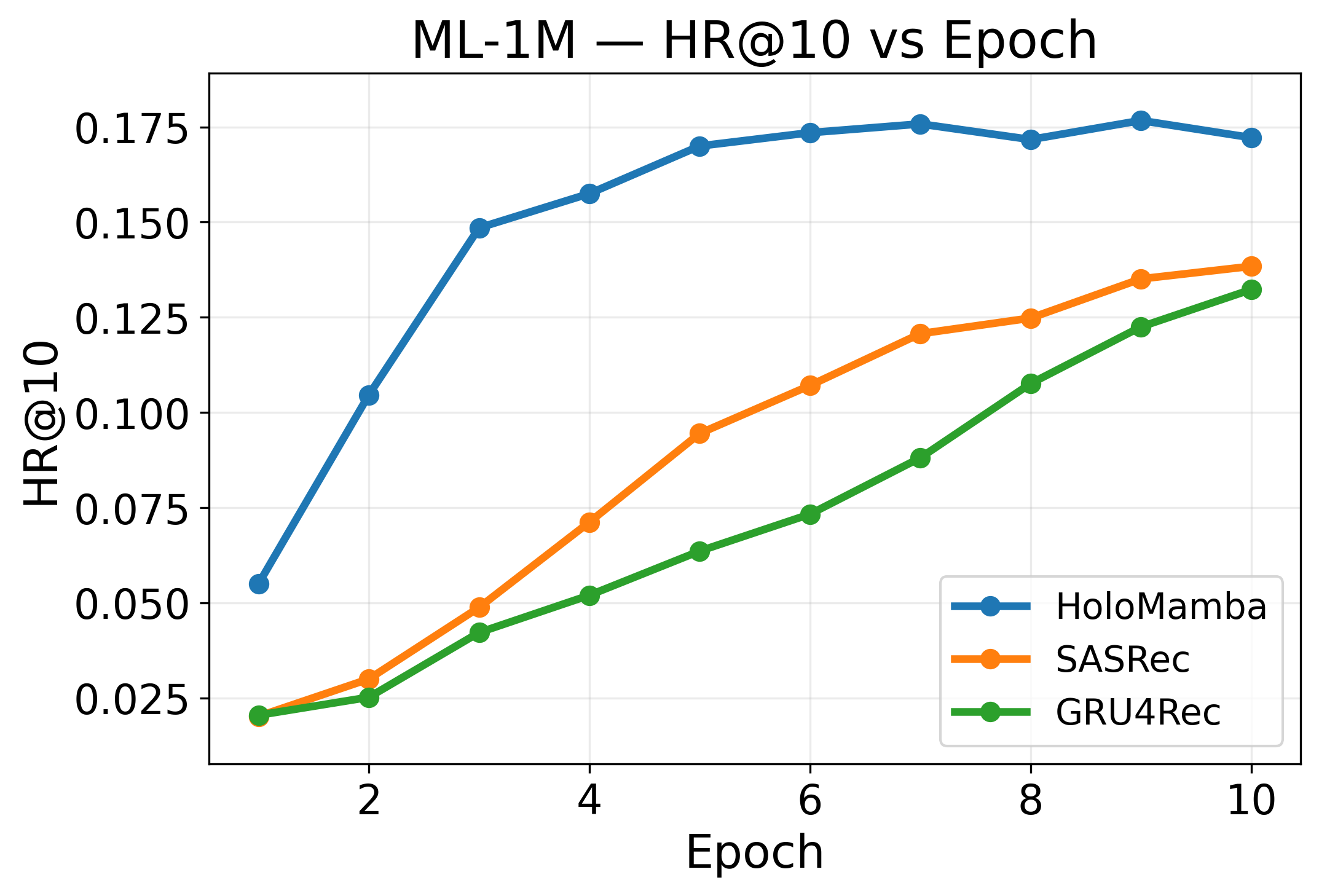} \\[4pt]
\includegraphics[width=0.48\columnwidth]{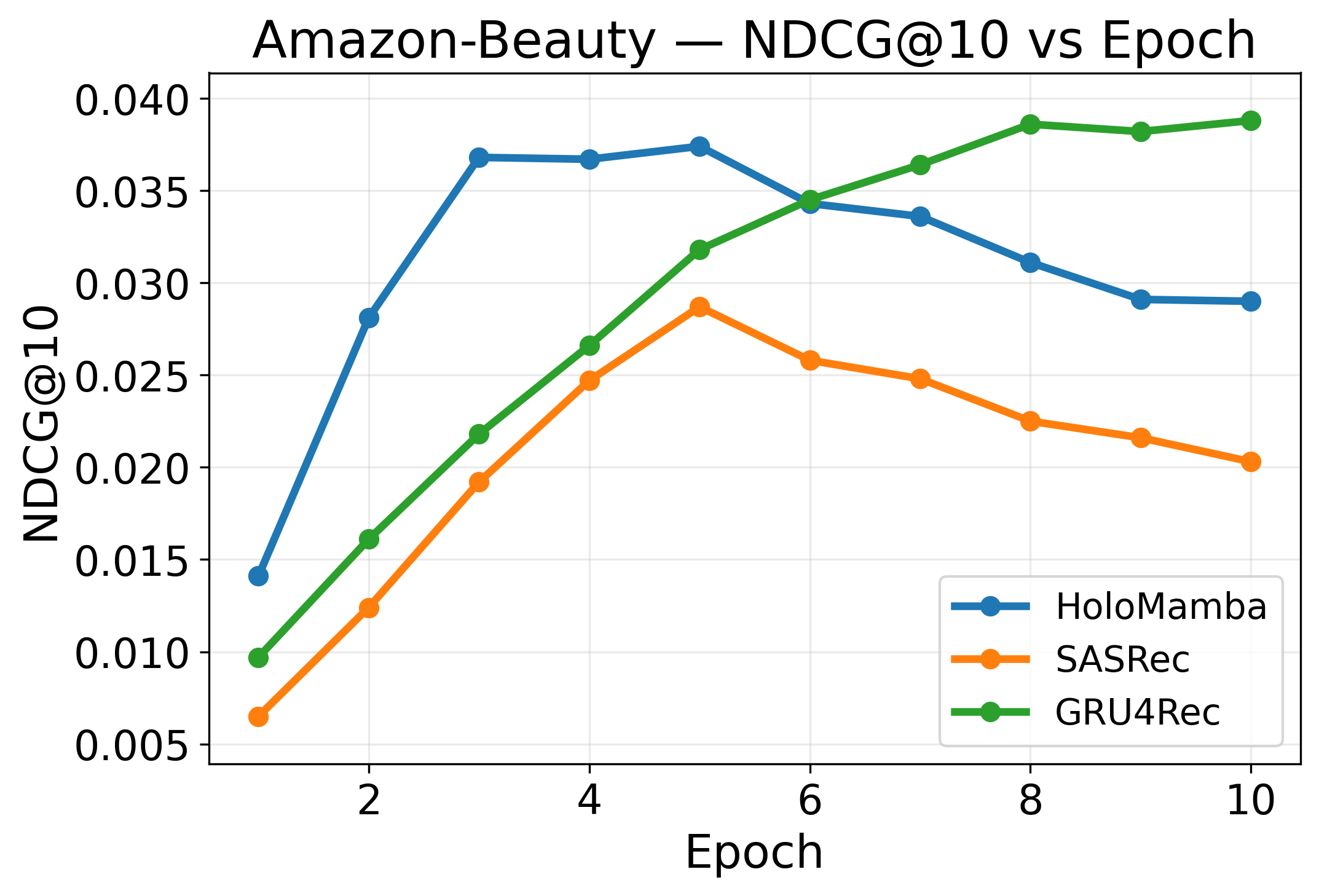} &
\includegraphics[width=0.48\columnwidth]{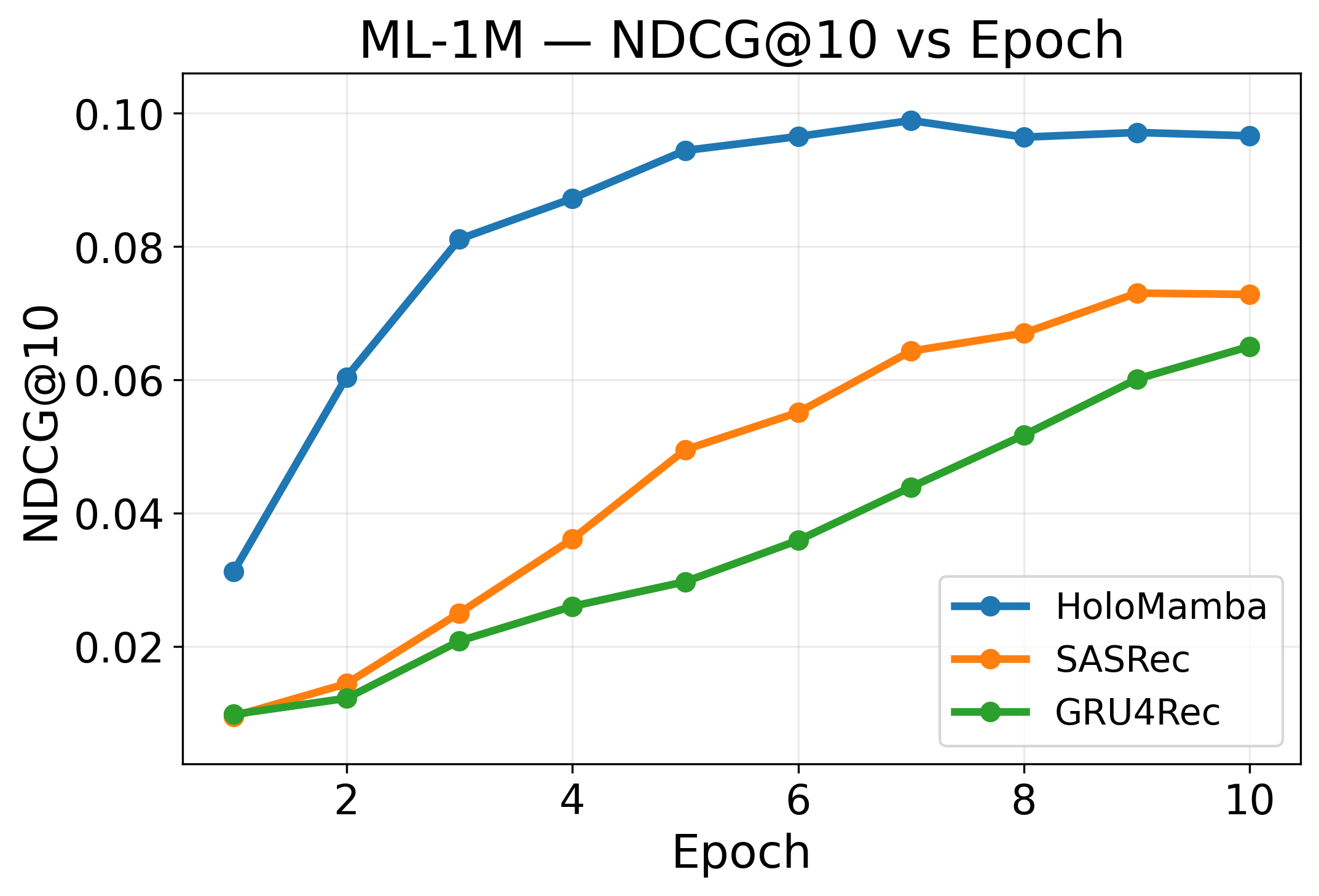}
\end{tabular}
\caption{Hit Rate and NDCG learning curves on Amazon Beauty and MovieLens-1M}
\label{fig:bench}
\end{figure}

Fig.~\ref{fig:bench} reports learning curves for HR@10 and NDCG@10 across epochs. HoloMambaRec exhibits stable and monotonic improvements during training, converging faster than SASRec on both datasets. On MovieLens-1M, HoloMambaRec maintains a consistent advantage throughout training, whereas on Amazon Beauty the gap between HoloMambaRec and GRU4Rec narrows but does not fully close within the 10-epoch budget. This suggests that longer training schedules or richer side information may further benefit the proposed architecture.

\subsection{Compression analysis: optimization paradox (negative result)}
We activated holographic bundling ($k{=}4$) and trained with \texttt{use\_compression=True}. Despite loss dropping sharply (Amazon 8.92$\rightarrow$0.80; ML-1M 7.44$\rightarrow$1.37), HR/NDCG collapsed to 0.0055/0.0027 (Amazon) and 0.0126/0.0060 (ML-1M). When training uncompressed but evaluating with compression for long sequences ($L{=}512$), accuracy again collapsed (Amazon HR 0.00067; ML-1M HR 0.00182) even as latency improved (Amazon 102.6$\rightarrow$41.7 ms; ML-1M 185.5$\rightarrow$43.7 ms) and peak VRAM dropped modestly (ML-1M 1492$\rightarrow$968 MB; Amazon roughly flat). This ``optimization paradox'' shows the model learns bundled representations but the decoder cannot recover ranked items; compression remains an open research problem and is reported as a negative result.

\subsection{Ablation: binding vs.\ item-only}
To isolate the effect of holographic binding, Table~\ref{tab:ablate} reports an ablation comparing HoloMambaRec against an item-only Mamba encoder in which attribute embeddings and binding operations are removed. Both variants are trained for 10 epochs under a shared configuration with $n_{\mathrm{layers}}{=}2$ and batch size $128$ on both datasets.

\begin{table}[H]
\centering
\caption{Ablation on holographic binding}
\label{tab:ablate}
\setlength{\tabcolsep}{6pt}%
\begin{tabular}{lcccc}
\toprule
Model & \multicolumn{2}{c}{Amazon Beauty} & \multicolumn{2}{c}{ML-1M} \\
\cmidrule(lr){2-3}\cmidrule(lr){4-5}
& HR@10 & NDCG@10 & HR@10 & NDCG@10 \\
\midrule
HoloMambaRec (binding) & 0.0431 & 0.0268 & 0.1695 & 0.0945 \\
Mamba (item-only)   & 0.0424 & 0.0268 & 0.1728 & 0.0976 \\
\bottomrule
\end{tabular}
\end{table}

The ablation indicates that holographic binding provides limited benefit under the current experimental setup. On Amazon Beauty, binding yields near-identical performance to the item-only variant, while on MovieLens-1M it slightly trails in both HR@10 and NDCG@10. This outcome is consistent with the use of a single synthetic or coarse-grained attribute, which provides limited additional signal beyond item identity. The results suggest that the full potential of holographic binding is likely to emerge when richer metadata or multiple attributes per item are incorporated, or when temporal bundling is jointly trained to exploit superposition and compression effects.

\section{Practical Notes}
Circular convolution enables attribute-aware item representations while preserving the original embedding dimensionality, ensuring that the encoder’s parameter count and memory footprint remain fixed as additional metadata is introduced. The selective state space scan is executed explicitly over timesteps, which makes the recurrent dynamics transparent and interpretable while retaining linear $\mathcal{O}(L)$ computational complexity. All experiments are conducted using a consistent and reproducible configuration with embedding dimension $d=96$, state size $d_{\mathrm{state}}=16$, and $n_{\mathrm{layers}}{=}2$ for Amazon Beauty and $3$ for MovieLens-1M, optimized with AdamW at a learning rate of $10^{-3}$, batch size $64$, and a fixed training budget of $10$ epochs. Temporal compression via holographic bundling is evaluated separately as a negative result: it improves latency and memory but destroys ranking accuracy; thus it remains future work rather than part of the core claim.

\section{Limitations and Future Work}
This study focuses on establishing feasibility and is intentionally limited in scope. The current design binds only a single discrete attribute per item, whereas production recommender systems typically rely on richer and multi-valued metadata. Extending holographic binding to multiple attributes and analyzing interference when several roles are superimposed are important directions for future work, as is understanding whether selective state space models can reliably recover individual roles from such superpositions.

Efficiency and evaluation also present limitations. While temporal bundling offers clear potential for reducing latency and VRAM, its accuracy trade-offs without joint training remain unresolved. Enabling bundling during training and adapting prediction targets to operate at the window level will be necessary to realize practical compression benefits. In addition, the selective scan is currently implemented in Python; future work should explore fused CUDA or Triton kernels and comparisons with optimized Transformer variants.

Finally, evaluation is restricted to HR@10 and NDCG@10 on two datasets with a short training schedule. Broader metrics, longer training horizons, larger catalogs, and deployment-oriented measures such as throughput, cold-start robustness, and online latency behavior will be required to fully characterize the strengths and limitations of HoloMambaRec.

\section{Conclusion}
This paper presents HoloMambaRec, a hybrid sequential recommendation architecture that integrates holographic reduced representations with a selective state space encoder to overcome the scalability and memory constraints of attention-based models. Item identifiers are bound with discrete attributes via circular convolution, and sequence dynamics are modeled using linear-time selective state spaces, enabling attribute-aware representations without increasing embedding dimensionality or incurring quadratic attention overhead.

Empirical evaluation on Amazon Beauty and MovieLens-1M under a constrained 10-epoch training budget shows that HoloMambaRec consistently outperforms SASRec on both datasets and achieves the strongest overall results on MovieLens-1M. Compared to GRU4Rec, the model remains competitive on sparse data and demonstrates clear advantages on denser interaction sequences, highlighting the effectiveness of selective state space dynamics for long-horizon modeling under practical resource constraints. Learning curves further indicate stable optimization behavior and faster convergence relative to Transformer baselines.

Ablation results suggest that single-attribute holographic binding provides limited gains in the current setting, underscoring that richer metadata or multi-attribute fusion is likely required to fully realize the benefits of holographic representations. Importantly, the architecture is designed to be forward-compatible: temporal bundling, multi-attribute binding, and optimized kernel implementations can be activated and trained end-to-end to further reduce latency and memory usage while expanding representational capacity.

Overall, HoloMambaRec constitutes a practical step toward scalable and deployment-ready sequential recommendation systems, enabling long-horizon user modeling, efficient integration of metadata, and operation within realistic latency and memory constraints.

\end{document}